\documentclass[11pt]{article}
\usepackage{amssymb}
\usepackage{epsfig}

\def\slash#1{\settowidth{\bredde}{$#1$}\ifmmode\,\raisebox{.15ex}{/}
\hspace*{-\bredde} #1\else$\,\raisebox{.15ex}{/}\hspace*{-\bredde} #1$\fi}

\newcommand {\beq} {\begin{equation}}
\newcommand {\eeq} {\end{equation}}
\newcommand {\beqa}{\begin{eqnarray}}
\newcommand {\eeqa}{\end{eqnarray}}
\newcommand {\dd}{\mbox{d}}
\newcommand {\tr}{{\rm tr\,}}
\newcommand {\Tr}{\mbox{Tr\,}}
\newcommand{\sect}[1]{\setcounter{equation}{0}\section{#1}}

\begin{document}

\topmargin -1.4cm
\oddsidemargin -0.8cm
\evensidemargin -0.8cm

\title{\Large{{\bf Rotational Symmetry Breaking in Multi-Matrix Models}}}

\vspace{1.5cm}

\author{~\\{\sc Graziano Vernizzi\footnote{vernizzi@thphys.ox.ac.uk}}
\\~\\and\\~\\
{\sc John F. Wheater\footnote{j.wheater1@physics.ox.ac.uk}}\\~\\
Theoretical Physics, Department of Physics, University of Oxford,\\ 1
Keble Road, OX1 3NP, Oxford, UK\\ }
\date{\today}

\maketitle

\begin{center}
{\it Dedicated to the memory of Jo\~ao D.~Correia} 
\end{center}
\vspace{1.5cm}

\begin{abstract}
\noindent
We consider a class of multi-matrix models with an action which is
$O(D)$-invariant, where $D$ is the number of $N \times N$ Hermitian
matrices $X_\mu$, $\mu=1,\ldots,D$. The action is a function of all
the elementary symmetric functions of the matrix $T_{\mu\nu}=Tr(X_\mu
X_\nu)/N$. We address the issue whether the $O(D)$ symmetry is
spontaneously broken when the size $N$ of the matrices goes to
infinity. The phase diagram in the space of the parameters of the
model reveals the existence of a critical boundary where the $O(D)$
symmetry is maximally broken.
\end{abstract}


\vfill

\begin{flushleft}
OUTP-02-29P
\end{flushleft}
\thispagestyle{empty}
\newpage

\renewcommand{\thefootnote}{\arabic{footnote}}
\setcounter{footnote}{0}

\sect{Introduction}
Over the last twenty years several multi-matrix models have been
considered for the description of a wide range of physical systems, from
Statistical Physics to QCD or Quantum Gravity
\cite{GMW,JT,DiFrancesco:1993nw,Ambbook}.  Although an analytic solution is
generally not as easy to achieve as for the single-matrix models, a
remarkable number of successes and results have been obtained so far 
\cite{Kazakov:2000aq}. A general feature of one-matrix models is that
they possess an internal global symmetry under some gauge group
(e.g. $U(N)$ invariance, where $N$ is the size of the matrix) which
determines much of the universal behaviour in the large $N$
limit. This global symmetry is present also in all the most relevant
multi-matrix models (Ising model on random lattice
\cite{Kazakov:ising,BK}, the $Q$-state Potts model 
\cite{Kazakov:NPB,Kostov:NPB,Daul:1994qy,Zinn-Justin:1999jg,Bonnet:1999nf,Eynard:1999gp},
chain of matrices \cite{Kostov:1992ie, Kharchev:1992iv, Eynard:1998,Eynard:1997jf,Eynard:199,Chadha:1980ri}, models
for coloring problem
\cite{Cicuta:1992ue,DiFrancesco:1997fp,Chekhov:1996xy,Eynard:fn,Kostov:2000vn},
vertex models
\cite{Zinn-Justin:1999wt,Kostov:1999qx,Johnston:1997vy,Johnston:1999az,Kazakov:1998qw},
the meander model \cite{DiFrancesco:1995cb,Makeenko:1995sa}, the
$O(n)$-model and some generalizations of it
\cite{Gaudin:vx,Kostov:fy,Kostov:pn,Eynard:1992cn,Eynard:1995nv,Eynard:1995zv,Durhuus:1996gm,ZJ99,ZJ20,ZJbm}, and several others
\cite{Engelhardt:1996da,Eynard:1998kg,DiFrancesco:1999pn,Itzykson:1979fi,Mehta:xt,Staudacher:1993xy,Douglas:1994av,Daul:1993bg,Bertola:2001br}. The
list is not complete). However, they do not usually have any further
symmetry, except for the $O(n)$ model and its generalizations where the
whole set of matrices transform as a $O(n)$ vector. The symmetry of
these models is then $U(N) \times O(n)$. Recently a new  class of
multi-matrix models have been introduced in the
framework of Superstring Theory and M-theory and the two main
representative are the so-called IKKT-model \cite{IKKT1,IKKT2,IKKT3} and the BFSS
model \cite{BFSS}. They are proposed  to be
a non-perturbative definition of type IIB Superstring theory and
M-Theory, respectively.  In particular the IKKT-model is just one
element of a bigger class of matrix models, called Super Yang-Mills
Integrals (for an introduction see \cite{SYM,WA1,WA2,PA}). The latter are
characterized by carrying several (super)symmetries and they are
obtained from the complete dimensional reduction of $D$-dimensional
$SU(N)$ Super Yang-Mills theories.  These integrals also might provide
an effective tool for the calculation of the bulk Witten index of a
supersymmetric quantum mechanics theory \cite{Windex1,Windex2,Windex3,Windex4,Windex5}.

One consequence of having several symmetries is the existence of
flat directions in the action of the model. They are potential
sources of divergences when evaluating the integrals. The precise
domain of existence of all the  Yang-Mills Integrals with and without
supersymmetry and for all the gauge groups, has been rigorously
determined in \cite{WA1,WA2,PA} (after numerical and analytical studies for
small gauge groups in \cite{smallN1,smallN2,smallN3,smallN4,smallN5}, and for large gauge groups in
\cite{largeN11,largeN12,largeN13,largeN14,largeN21,largeN22,Moore}). The existence of such flat directions affects
 not only the convergence properties of the Super Yang-Mills
 integrals, but also the behaviour of all the correlation functions
 and of the spectral density asymptotics. During the last few years it has been claimed that the ``rotational''
$O(D)$ symmetry (where $D$ is the number of matrices) might be
spontaneously broken in the large $N$ limit \cite{AIKKT}.  This issue
has been analyzed in a series of analytical and numerical studies
\cite{largeN11,largeN12,largeN13,largeN14,largeN21,largeN22,NV,NS,N,Kawai:2002jk,recentNS} and a possible mechanism
for having such a spontaneous symmetry breaking has been proposed in
\cite{NV,NS,N}.

The basic idea relies on the fact that these integrals contain
fermionic degrees of freedom (i.e. matrices with Grassmannian entries)
in such a way that the action is a complex number in general. However
the action is a real number for lower-dimensional ``degenerate''
configurations (i.e. when the matrices are linearly
dependent). Therefore, when summing over all possible configurations
in the partition function the rapid oscillations of the complex action
might enhance lower dimensional configurations in the large $N$
limit. In order to shed light on this mechanism, a class of simplified
fermionic multi-matrix models having a complex action (and the same
$O(D) \times U(N)$ symmetry) has been studied in \cite{N}. In that
case, the symmetry breaking actually occurs, and it is shown to be a
consequence of the fact that the action is complex. Also, the results
in \cite{Kawai:2002jk} give indications of a spontaneous symmetry
breaking in the IKKT-model. However, the actual mechanism for having
such a behaviour (if confirmed) remains an open question.
 
In this paper we address the question of whether a {\em complex}
action is necessary if there is to be a spontaneous breaking of the
$O(D)$ symmetry at large $N$. The action of the Super-Yang Mills
Integrals is complex in general but it also has flat directions. These
two features have a quite different origin. The former is a
consequence of the particular choice of the structure of the spinors
together with the signature of the $D$-dimensional ``space-time'' in
consideration. The latter arises because the action is made up of
commutators or logarithms of fermionic determinants (or
Pfaffians) (which are there ultimately as a consequence of having an
highly symmetric theory).  Since it happens that along the flat
directions the action becomes {\em real}, it is not clear whether the
spontaneous symmetry breaking is a consequence of the complexity of
the action, or its flatness properties. A definite answer to this
question would be given by a complete analytic solution of real-action models
 such as the Super Yang-Mills integral in four dimension, or its 
bosonic version (at any $D$) in which the fermions are
suppressed. However only numerical simulations are available so far.
The results of \cite{Hotta:1998en} suggest that there is no
spontaneous symmetry breaking in the pure bosonic Yang-Mills integral.
About the $4D$ Super Yang-Mills integral there has been some dispute  
\cite{largeN14,largeN22} whether there is symmetry breaking or not, and 
about which is the most reliable order parameter to use in that case 
 (for a review see \cite{review1,review2}).

We decide then to focus our attention on building-up a multi-matrix
model with a real positive semi-definite action made of standard
Hermitian matrices (``bosonic''), but which allow a wide class of
possible ``degenerate'' configurations. In this paper, we shall
introduce a multi-matrix model sharing the same $O(D) \times U(N)$
symmetries, but with a real positive weights and without any
Grassmannian degrees of freedom. This action allows many degenerate
configurations and we will find that they can affect the symmetry  
of the model at large $N$.  This fact is an indication
that the exact mechanism which could be at the origin of a possible
spontaneous symmetry breaking of rotational symmetries in
Super Yang-Mills integrals deserves further studies.

The paper is organized as follows: in Section \ref{modelSec} we define
our multi-matrix model. It is based on all the elementary symmetric
functions of the eigenvalues of the two-point correlation matrix, and
it is manifestly $O(D)$ invariant at finite $N$. The model contains a
number of coupling constants which controls the r\^ole of the various
elementary symmetric functions and the interaction among them. We
study the behaviour of the model in the space of such parameters.  In
particular we solve the model in the simple and illuminating case
where only two basic elementary symmetric functions are involved,
i.e. the trace and the determinant.  This case is simple enough for
carrying explicit calculations at large $N$ by means of a saddle-point
method.  In Section \ref{genSec} we consider the more general case
where all the elementary symmetric functions are present. There we
show how the model is stable under such a generalization and that the
$O(D)$ symmetry of the system holds everywhere except on a critical
boundary where the symmetry is maximally broken.  Finally, Section
\ref{concSec} is devoted to our discussions and conclusions. For the
sake of completeness, the appendix contains the calculation of a
Jacobian we make use in Section \ref{modelSec}.


\sect{The model} \label{modelSec}
Let us consider a set of $N \times N$ Hermitian matrices $\{ X_\mu \}$, 
$\mu=1,\ldots,D$.  The corresponding two-point correlation matrix
\beq
\label{matrixT}
T_{\mu\nu} \equiv \frac{1}{N} \Tr (X_\mu X_\nu) 
\eeq
is a $D \times D$ real symmetric positive semi-definite matrix, with
eigenvalues $ t_1 \ge \cdots \ge t_{D} \ge 0$.  From the
definition~(\ref{matrixT}) we see that if $X_{\mu} \to X_{\mu}'=Q_{\mu\nu}
X_{\nu}$ where $Q \in O(D)$, then $T$ transforms as $T'=QTQ^{T}$. A
straightforward consequence is that all the eigenvalues $t_\mu$ of $T$
are $O(D)$ invariant quantities.  Moreover, the matrices $\{ X_\mu \}$ 
are linearly dependent iff some of the eigenvalues $t_\mu$ of the
correlation matrix $T_{\mu\nu}$ are identically zero\footnote{A short
proof: if $\{X_\mu \}$ are linearly dependent, then $\exists
\eta_\mu$ not all zero such that $\sum_\mu \eta_\mu X_\mu=0$. Therefore $\sum_\nu
T_{\mu \nu} \eta_\nu=0$, i.e. $T$ has a zero eigenvalue. On the other
hand, if $\sum_\nu T_{\mu \nu} \eta_\nu=0$, then $\tr [(\sum_\mu X_\mu
\eta_\mu)^2]=\sum_{\mu,\nu} \eta_\mu T_{\mu \nu} \eta_\nu=0$ which implies 
$\sum_\mu X_\mu \eta_\mu=0$.}. More
precisely, a good indicator of the degree of non-degeneracy of the
matrices $\{ X_\mu \} $ is $r(T) \equiv rank(T)$, i.e. the number of
non-zero eigenvalues of the matrix $T$. The most general action
which is $O(D)$ invariant and is a function of the variables $t_\mu$
only, can be expressed in terms of the elementary symmetric functions
$c_k$ of the variables $\{t_\mu \}$, $k=0, \ldots, D$. We recall here
that the $k$-th order elementary symmetric function of the variables
$\{t_\mu\}$ is defined as the products of $k$ distinct variables $t_\mu$
\beq
\label{c_definition}
c_k = \sum_{\mu_1<\mu_2<\ldots<\mu_k} t_{\mu_1} t_{\mu_2} \ldots t_{\mu_k} \quad .
\eeq
(we omit the explicit $t_\mu$-dependence of $c_k$).  It is well known
that the $c_k$ can be obtained from
the expansion of the characteristic polynomial of the matrix $T$
\beq
\label{gen_fun}
C(z) \equiv \det({\mathbb I}_{D \times D}+ z T)= c_D z^D+c_{D-1} z^{D-1}+\cdots+ c_0 \, .
\eeq
All the $c_k$ are non-negative, as the matrix $T$ is
positive semi-definite.  In particular one has 
\[
c_1= \tr  T = \frac{1}{N} \sum_{\mu=1}^{D} \Tr (X_{\mu}^2) \, , \quad   
c_D= \det \left(  \frac{1}{N} \Tr X_{\mu} X_{\nu} \right) \, ,
\]
where we use the symbol ``Tr'' and ``tr'' to indicate the trace
over $N \times N$ and $ D\times D$ matrices, respectively.\\

The partition function we consider in this paper is
\beq
\label{Z}
{\cal Z}[\alpha] =  \int \prod_{\mu=1}^D \dd X_\mu \, e^{- N  \Tr \sum_{\mu=1}^D X_{\mu}^2  }
\prod_{k=1}^{D} \left( c_k  \right)^{\alpha_k N^2} \, ,
\eeq
where $\alpha_k$ are real parameters.  Eq.~(\ref{Z}) is manifestly
$O(D)$ invariant. This symmetry is not to be confused with the usual
$U(N)$ ``internal'' symmetry, which still holds for this model. In
fact ${\cal Z}[\alpha]$ is invariant under $X_\mu \rightarrow U X_\mu
U^{\dagger}$, for all $\mu$, with $U \in U(N)$. The region of
existence of this model as a function of the real parameters
$\alpha_k$ will be determined later in this Section. Here we just
emphasize that the argument of the matrix-integrals is always real and
positive semi-definite. Moreover, another feature of eq. (\ref{Z}) is
the existence of ``flat directions''. They correspond to
configurations where the matrices $\{ X_\mu \}$ are linearly
dependent, i.e. such that some of the symmetric functions $c_k$ are
identically zero. The convergence properties of the integral (\ref{Z})
for large values of the entries of the matrices are mainly guaranteed
by the presence of the Gaussian weight, but not completely. In fact,
the flat directions contain non-integrable singularities
(with some analogy to the case of the Yang-Mills integrals
\cite{WA1,WA2,PA}) when some of the parameters $\alpha_k$ are too
negative. An exact bound in the space of the parameters $\{ \alpha_k
\}$ for the existence of eq.  (\ref{Z}) is presented in 
eq. (\ref{condD_largeN}). At finite $N$ the average eigenvalues $
\langle t_\mu \rangle$ of the matrix $T$ are all equal, because of the
$O(D)$ invariance of eq.~(\ref{Z}).  However at large $N$ this may no
longer be the case, and our aim is to see whether the $O(D)$
rotational symmetry of the model can be spontaneously broken when $N
\rightarrow \infty$.  In this context, we define also the {\it
dimensionality} $d$ of a configuration of matrices $ \{ X_\mu \}$ as
the number of non-vanishing eigenvalues of the average correlation
matrix $\langle T \rangle$.  Of course, at finite $N$ one always has
$d=D$. A possible way for probing $O(D)$ symmetry breaking is to
introduce an explicit symmetry breaking term before taking the large
$N$ limit. We do this by modifying the Gaussian weight in
eq. (\ref{Z}) $ e^{-N
\sum_{\mu}  \Tr (X_{\mu}^2) } \to e^{-N
\sum_{\mu} \lambda_{\mu} \Tr (X_{\mu}^2) } $, where the variables
$0<\lambda_1 < \lambda_2 < \cdots < \lambda_D$ maximally break the
$O(D)$ symmetry of the model (in analogy to \cite{N}).  After taking
the large $N$ limit, we shall remove the symmetry breaking term by
taking the limit $\lambda_\mu \to 1, \forall \mu$. If $ \langle t_\mu
\rangle \to 0$ 
for different directions $\mu$ then there is spontaneous symmetry
breaking of the $O(D)$ symmetry.\\

\begin{figure}[h!]
\centerline{
\epsfig{figure=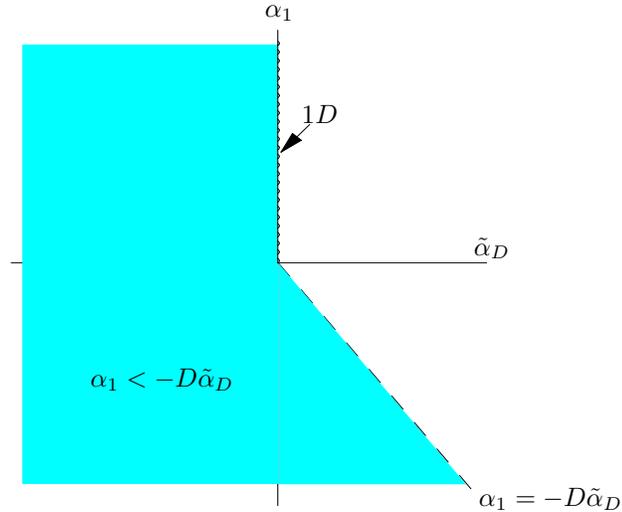,height=15pc,width=15pc,clip=}
\put(-84,185){\small $\alpha_1$}
\put(-5,95){\small $\tilde{\alpha}_D$}
\put(-3,0){\small $\alpha_1=-D \tilde{\alpha}_D$}
\put(-150,45){\small $\alpha_1 < -D \tilde{\alpha}_D$}
\put(-70,145){\small $1D$}
}
\caption{Phase diagram of the model in eq. (\ref{Zsimple}). The shaded 
region is where the partition function is divergent. The wiggle line
is the region where the model is one-dimensional. In the remaining region 
 the model maintains the full $O(D)$ dimensionality.}
\label{fig1}
\end{figure}

We start with the simple case where
$\alpha_2=\alpha_3=\cdots=\alpha_{D-1}=0$. The partition function
reads
\beq
{\cal Z}[\alpha,\Lambda]= \int \prod_{\mu=1}^D \dd
 X_\mu \, e^{- N \Tr \sum_{\mu=1}^D \lambda_\mu X_{\mu}^2 } \left(\tr T
\right)^{\alpha_1 N^2} \left(\det T \right)^{\alpha_D N^2}
\label{Zsimple}
\eeq
It is convenient to introduce the matrix $\Lambda=\delta_{\mu \nu}
\lambda_\mu$, so that the partition function can be written as $
{\cal Z}[\alpha,\Lambda]= \int \prod_{\mu=1}^D \dd X_\mu \, \exp( - N^2
S_0 )$ where the action is
\beq
\label{S}
S_0[T,\alpha,\Lambda]= \tr ( \Lambda T) - \alpha_1 \log \tr T  - \alpha_D \log \det T \, .
\eeq
The action $S_0$ depends on all the matrices $X_\mu$ only through the
matrix $T$: it is therefore natural to change the integration measure
from the multi-matrix variables $\{ X_\mu \}$ to the single-matrix $\{
T_{\mu \nu} \} $.  When $N^2 \geq D$ we have (see the Appendix
\ref{Appendix} for details)
\beq
\label{ZT}
{\cal Z}[\alpha,\Lambda] = {\cal C}_{N,D} \int_{T \geq 0} \dd T 
\,  e^{- N^2 S[T,\alpha,\Lambda]+\frac{N^2-D-1}{2}\log (\det T)  }
\, , \quad {\cal C}_{N,D}= \frac{N^{\frac{D N^2}{2}}\pi^{\frac{D}{4}
(2N^2-D+1)}}{2^{D\frac{N(N-1)}{2}}
\prod_{k=1}^D \Gamma \left( \frac{N^2-k+1}{2} \right) } \, , 
\eeq
where the integral is over all the $D \times D$ real symmetric
positive-definite matrices, the measure is $d T= \prod_{\mu \geq \nu}
dT_{\mu\nu}$ and the Jacobian of the transformation is proportional to
$\det(T)^{\frac{N^2-D-1}{2}}$. The partition function now has the
proper form for the study of the large $N$ limit by means of the
saddle-point (Laplace) method for the asymptotic expansions of
multidimensional integrals.  According to this method, the main
contribution to the integral comes from a small neighborhood of the
critical points, i.e. global minimum points in this case, of the 
action (we drop $1/N^2$ sub-leading terms)
\beq
S[T,\alpha,\Lambda]\equiv 
\tr T \Lambda -  \alpha_1 \log \tr T  -\tilde{\alpha}_D \log \det T  \, ,
\label{Seff}
\eeq
where $\tilde{\alpha}_D \equiv \alpha_D+1/2$.
The minima of the function $S$ can be at the boundary of the
integration region or at the interior of it. In the latter case the
necessary stationarity conditions for having a minimum are
({\it saddle-point equations})
\beq
\label{saddle}
\frac{\partial}{\partial_{T_{\mu \geq \nu}}} S[T,\alpha,\Lambda]= \lambda_{\mu} \delta_{\mu\nu}-
\alpha_1 \frac{\delta_{\mu \nu}}{\tr T} -\tilde{\alpha}_D(T^{-1})_{\nu\mu}=0 \, ,
\eeq
for all $1 \leq \nu \leq \mu \leq D$. Note that multiplying
eq.~(\ref{saddle}) by $T$ and taking the trace gives $ \tr (T
\Lambda)=\alpha_1+D \tilde{\alpha}_D$. Since $\tr (T
\Lambda) \geq 0$  we have to look for solutions of
eq. (\ref{saddle}) in the region of the parameters plane
$\{\alpha_1,\tilde{\alpha}_D \}$
\beq
\alpha_1 \geq -D  \tilde{\alpha}_D \, .
\label{domain}
\eeq
The condition (\ref{domain}) is actually a bound on the domain of
existence of the model at large $N$. In fact as we have already
announced, the integral in eq. (\ref{ZT}) exists only when the
parameters $\alpha_k$ satisfy suitable constraints, and
eq. (\ref{domain}) is one of them.  Namely, the integrand function in
eq. (\ref{ZT}) does not have singularities in the integration region,
except perhaps at the integration boundaries.  At large values of the
entries of $T$ the integrand function is regular and integrable for
any value of $\alpha_k$, being bounded by the exponential factor.
However, the behaviour close to the origin can give non-integrable
singularities. This fact is evident when passing to the eigenvalues
$\{ t_\mu \} $ of $T=O t O^{T}$. It yields
\beq
\label{finiteZ}
{\cal Z}[\alpha,\Lambda] \sim \int_0^{\infty} \prod_{\mu=1}^{D} \dd t_\mu 
\left| \Delta(t) \right|
\,  
\int_{O(D)} \dd O \,
e^{- N^2 \tr O t O^{T} \Lambda} 
\left( \sum_{\mu=1}^{D} t_\mu \right)^{\alpha_1 N^2}
\left( \prod_{\mu=1}^{D} t_\mu \right) ^{\alpha_D N^2+\frac{N^2-D-1}{2}} \, ,
\eeq
where $\Delta(t)$ is the Vandermonde determinant $\prod_{\mu<\nu}^{D}
(t_\mu-t_\nu)$, $\int_{O(D)}$ is the integral over $D \times D$
orthogonal matrices, with Haar measure $\dd O$, $t$ is a diagonal
matrix with diagonal elements $t_1,\ldots,t_D$ and ``$\sim$'' means
``up to a (irrelevant) proportionality constant''.  From
eq. (\ref{finiteZ}) we see that, first, in order to have an integrable
singularity at each of the $(D-1)$-dimensional boundary where only one $t_\mu=0$, 
it has to be
\beq
\alpha_D N^2+\frac{N^2-D-1}{2} > -1 \, .
\eeq
At large $N$ this condition simplifies to $\tilde{\alpha}_D=\alpha_D+1/2
\geq 0$. Secondly, by rewriting the integral in eq. (\ref{finiteZ})
from cartesian coordinates into multi-dimensional spherical
coordinates, one has that the radial integration exists if and only if
\beq 
(D-1)+\frac{D(D-1)}{2} + N^2 \alpha_1 + \left( \frac{N^2-D-1}{2}+N^2 \alpha_D \right) D > -1 \, .
\label{usual}
\eeq
Note that there are no contributions from the integral over the
orthogonal group: in fact it is  finite and regular in $t$, since it is an
integral over a compact domain of an analytic function in its variables. 
At large $N$ the condition (\ref{usual}) is fulfilled by
$\alpha_1+(\frac{1}{2}+\alpha_D) D \geq 0$ which is precisely
eq. (\ref{domain}). In summary, the region of existence of the model at large $N$ is 
\beq
\label{domaintrue}
{\cal D}
\equiv \left\{ \{\alpha_k \} : \alpha_1+D \tilde{\alpha}_D \geq 0
\; \mathrm{and} \; 
\tilde{\alpha}_D \geq 0 \right\} \, ,
\eeq
and it is depicted in figure \ref{fig1}. We point out that the model
at large $N$ is well-defined and finite also on the boundaries of 
${\cal D}$,
i.e. ${\cal B}_1 \equiv \{\tilde{\alpha}_D =0, \alpha_1 > 0 \}$ and 
${\cal B}_0 \equiv \{\alpha_1=-D\tilde{\alpha}_D , \tilde{\alpha}_D \geq 0 \}$.

\begin{figure}[h!]
\centerline{
\label{fig2}
\epsfig{figure=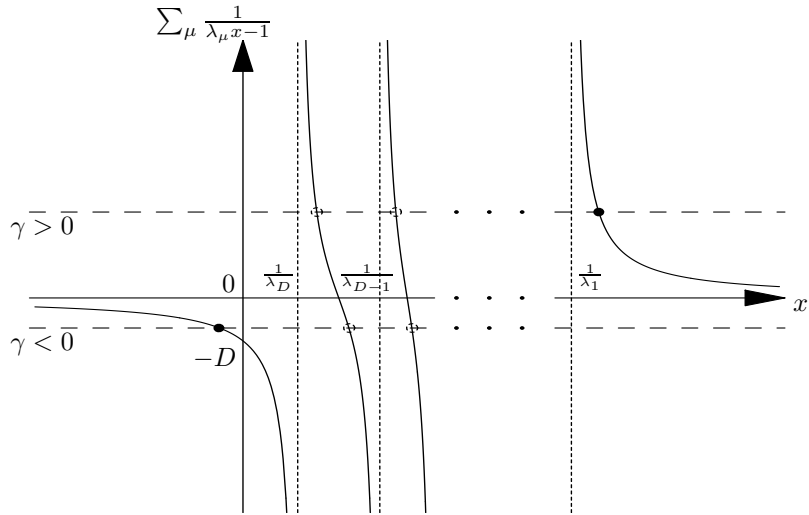,height=15pc,width=25pc,clip=}
\put(-300,106){\small $\gamma>0$}
\put(-300,62){\small $\gamma<0$}
\put(-231,56){\small $-D$}
\put(-220,84){\small $0$}
\put(-205,89){\tiny $\frac{1}{\lambda_D}$}
\put(-176,89){\tiny $\frac{1}{\lambda_{D-1}}$}
\put(-86,89){\tiny $\frac{1}{\lambda_1}$}
\put(-4,77){\small $x$}
\put(-245,184){\small $\sum_\mu \frac{1}{\lambda_\mu x -1}$}
}
\caption{
Graphical representation of the r.h.s. of eq. (\ref{eq4x}) as a
function of $x=\frac{\tr T}{\alpha_1}$.  The circles indicate all the
solutions. The black circles indicate the acceptable solutions.}
\end{figure}

If $\tilde{\alpha}_D>0$ then we see immediately that the global minima of
$S$ in eq. (\ref{Seff}) cannot be on the boundary of the
integration region. Otherwise the matrix $T$ would have at least a
zero eigenvalue, that is $\det T=0$, and there  eq. (\ref{Seff}) gives
 $S \to +\infty$. Therefore,
in this case the critical points must be in the interior of the
integration region.  Let us then solve eq. (\ref{saddle}) for
$\tilde{\alpha}_D>0$.  It is straightforward to see that any matrix $T$
which is a solution of eq. (\ref{saddle}) has to be 
diagonal. Defining $T=
\delta_{\mu \nu} t_\mu$ eq. (\ref{saddle}) reads
\beq
 t_\mu=\tilde{\alpha}_D \left( \lambda_\mu-\frac{\alpha_1}{\tr T}
 \right)^{-1} \, , \quad \mu=1,\ldots D \, .
\label{sol_saddle}
\eeq
This system of algebraic equations can be solved easily. First, by
summing eq. (\ref{sol_saddle}) over $\mu$ we get an equation for $x
\equiv \tr T/\alpha_1$,
\beq
\gamma \equiv \frac{\alpha_1}{\tilde{\alpha}_D}=
\sum_{\mu=1}^{D}\frac{1}{\lambda_\mu x -1} \, , \quad x \equiv \frac{\tr T}{\alpha_1} \, .
\label{eq4x}
\eeq
For any given real $\gamma$ and $\Lambda$ eq. (\ref{eq4x}) is a
rational algebraic equation with $D$ solutions in the variable
$x$. All the solutions are real. In fact, by writing the real and
imaginary part of $x=x'+i x''$ and using the fact that $\gamma,\Lambda
\in \mathbf{R}$, it yields $x''=0$. Among such $D$ real solutions, we
have to pick up the ones that make $t_\mu \geq 0$ because $T$ has to
be a positive semi-definite matrix.  From eq. (\ref{sol_saddle}) we
obtain that
\beq
\left\{
\begin{array}{lcl}
x=\frac{\tr T}{\alpha_1} > \frac{1}{\lambda_1} & {\rm for} & \alpha_1>0 \\
\ & \ \\
x=\frac{\tr T}{\alpha_1} < \frac{1}{\lambda_D} & {\rm for} & \alpha_1<0 
\end{array}
\right. \, ,
\label{discussion2}
\eeq
which is satisfied by only one solution in each case. Namely, for
$\alpha_1>0$ is $\gamma>0$ and the solution is the largest possible
one (the one greater than $1/\lambda_1$) whereas for $\alpha_1<0$ is
$\gamma<0$ and the solution is the one with $x<0$  (see figure
\ref{fig2}).

For any given point $\{ \alpha_1, \alpha_D > 0 \}$ in the interior of
the parameter space ${\cal D}$, if we remove the symmetry breaking
terms  by taking $\lambda_\mu
\to 1 \, \forall \mu$,  then the (unique) solution of eq. (\ref{eq4x}) is 
$x=1+D/\gamma$, i.e. $\tr T= \alpha_1+\tilde{\alpha}_D D$. Inserting
this value in eq.  (\ref{sol_saddle}) we read that in the large $N$ limit 
all the eigenvalues are equal to $\langle t_\mu \rangle = \alpha_1/D+\tilde{\alpha}_D$. 
Hence, we conclude that in the region 
inside ${\cal D}$ with $\tilde{\alpha}_D \neq 0$ the model has a phase with
dimensionality $d=D$ and the $O(D)$ symmetry is preserved, as
expected. In such a phase, the free energy ${\cal F}=-\frac{1}{N^2}\log Z$ 
reads (with all $\lambda_\mu=1$)
\beq
\label{Fbulk}
{\cal F}=\left( \alpha_1 +D  \tilde{\alpha}_D \right)  
\left[ 1- \log \left( \alpha_1 +D  \tilde{\alpha}_D \right) 
\right]
+  D \tilde{\alpha}_D \log D \, .
\eeq

On the boundary ${\cal B}_0$ of ${\cal D}$ where
$\alpha_1=-D\tilde{\alpha}_D$ the solution of eqs. (\ref{sol_saddle})
and (\ref{eq4x}) gives $T = 0$. However, it would be wrong to conclude
that the dimensionality is $d=0$ there, because actually in the limit
$\tilde{\alpha}_D \to -\alpha_1/D$ one has $T \to 0$ with $d \to D$.
That means that there is no spontaneous symmetry breaking on the
boundary ${\cal B}_0$.

Let us consider now the final case of the boundary ${\cal B}_1$ where $\tilde{\alpha}_D
\to 0^{+}$ first and $\lambda_\mu \to 1$ for all $\mu$ afterwards. 
In such a limit one must consider $\alpha_1 >0$ in order to stay
within the region ${\cal D}$, eq. (\ref{domaintrue}), and therefore
$\gamma \to +
\infty$.\footnote{In principle it would be possible to take the same limit with $\alpha_1<0$ 
but then one necessarily would end up in the origin of the coordinates
$\alpha_1=\tilde{\alpha}_D=0$ where the system is purely Gaussian.} 
According to eq. (\ref{eq4x}) and figure
\ref{fig2}, this fact may occur only when $x \to 1/\lambda_1$. From
eq. (\ref{sol_saddle}) we have
\beq
\label{limitalpha}
\lim_{\tilde{\alpha}_D \to 0^{+}} \tr T =\frac{\alpha_1}{\lambda_{1}}  
\quad \textrm{and } \quad  t_{\mu}=\frac{\alpha_1}{\lambda_1} \delta_{\mu 1}  \, .\\
\eeq
In other words, only one eigenvalue of the matrix $T$ is not zero in
this limit.  Removing the symmetry breaking term by setting
$\lambda_{1} \to 1$ leads to a dimensionality $d=1$, actually $\langle
t_{\mu} \rangle=
\alpha_1 \delta_{\mu 1}$. That concludes our proof
that the model in eq. (\ref{Zsimple}) has a maximal spontaneous
symmetry breaking of $O(D)$ symmetry whenever $\tilde{\alpha}_D \to
0^{+}$.

It is interesting to notice that we could have had considered directly
the case $\tilde{\alpha}_D=0$ (and not just the limit
$\tilde{\alpha}_D \to 0^{+}$), because there the model at large $N$ is
well-defined. In fact, let $\tilde{\alpha}_D=0$ from the very
beginning in eq. (\ref{Seff}). Then, as the $\lambda_\mu$'s are all
different each other, $S$ cannot have any minima in the interior of
the integration domain (in other words, eq. (\ref{saddle}) do not
admit any solution). Hence, the global minimum must be on the
boundaries of the integration region, where some $t_\mu$ is equal to
zero. Analyzing by inspections all the hyperplanes which constitute
the integration boundary, one finds that the global minimum is a point
on the line $t_2=t_3=\ldots=t_D, t_1 >0$ and it is precisely at
$t_1=\alpha_1/\lambda_1$.  Substituting this value in eq. (\ref{Seff})
gives the free energy for the phase ${\cal B}_1$
\beq
\label{FB1}
{\cal F} =\alpha_1 (1- \log \frac{\alpha_1}{\lambda_1}) \, .
\eeq
This expression (for $\lambda=1$) matches continuously with the free
energy in the un-broken phase, eq. (\ref{Fbulk}) for $\tilde{\alpha}_D
\to 0^{+}$. By taking derivatives of the free energies with respect to
$\lambda_\mu$ we can compute the correlation functions, in particular
the average of the eigenvalues, and the susceptibility
\beq
\label{defder}
\langle t_\mu \rangle = \left. \frac{\partial {\cal F} }{\partial \lambda_\mu}\right|_{\lambda=1}  \, , \quad
\chi_{\mu \nu}=\left. 
\frac{\partial^2 {\cal F} }{\partial \lambda_\mu \partial \lambda_\nu} \right|_{\lambda=1} = -N^2 \langle t_\mu t_\nu \rangle_{conn} \, . 
\eeq
In the broken phase ${\cal B}_1$, we get from eq. (\ref{FB1}) and
(\ref{defder})
\beq
\label{B1}
\langle t_\mu \rangle = \alpha_1 \delta_{\mu 1} \, , \quad \chi_{\mu \nu}= -\alpha_1 \delta_{1 \mu} \delta_{1 \nu} \, .
\eeq
which is of course consistent with eq. (\ref{limitalpha}).  The
computation of the same quantities in the un-broken phase requires the
knowledge of an expression of the free energy as a function of
$\lambda_\mu$ (i.e. eq. (\ref{Fbulk}) is not useful for that). A
general analytic expression seems not so easy to get since it needs
the analytic solutions of the algebraic equation (\ref{eq4x}) in a
closed form, which is known to be an impossible task when the degree
of the equation is large. However, we can proceed as follows. We
already know the pattern of symmetry breaking from
eq. (\ref{limitalpha}). Hence we can restrict to the case where
$\lambda_1<\lambda_2=\cdots=\lambda_D$ without loosing in
generality. In this case, eq. (\ref{eq4x}) is a second order algebraic
equation which can be solved explicitly. We obtain then the free
energy, its first and second derivatives w.r.t. $\lambda_\mu$ and in
the limit $\lambda_\mu \to 1$ they are
\beq
\langle t_\mu \rangle= \frac{\alpha_1}{D}+\tilde{\alpha}_D \, ,  \quad
\chi_{\mu \nu}= 
\frac{1}{\tilde{\alpha}_D D^2} \left(
\frac{\alpha_1}{D}+\tilde{\alpha}_D \right) \times
\left\{
\begin{array}{ll}
-\alpha_1 (D-1)-\tilde{\alpha}_D D^2 & 
\mathrm{if } \; \mu=\nu \\
\alpha_1  & 
\mathrm{if } \; \mu \neq \nu 
\end{array}
\right.
\, .
\eeq
Note that the susceptibility is divergent as $\sim 1/\tilde{\alpha}_D$
when $\tilde{\alpha}_D \to 0$ . The singular behaviour of the
susceptibility is again a signal of a criticality at
$\tilde{\alpha}_D=0$, where the rotational symmetry is actually
maximally broken down to one dimension.

\sect{Generalization} \label{genSec}
Let us consider now the more general case eq. (\ref{Z}) where all the
symmetric functions are allowed (and not only $c_1$ and $c_D$,
i.e. the trace and the determinant, respectively). Again introducing
the symmetry breaking term $\Lambda_{\mu \nu}=\lambda_\mu \delta_{\mu
\nu}$, $0<\lambda_1<\lambda_2<\ldots<\lambda_D$, and following the
same path of reasoning as in the previous paragraph, we have
\beq
\label{finiteZ_gen}
{\cal Z}[\alpha,\Lambda] ={\cal C}_{N,D} \int_{T \geq 0} \dd T e^{-N^2
S_{N}[T,\alpha,\Lambda]} \, ,
\eeq
where the generalized action at finite $N$ is now:
\beq
S_{N}[T,\alpha,\Lambda] \equiv \tr T \Lambda - \sum_{k=1}^{D}
\tilde{\alpha}_k \log c_k +\frac{D+1}{2N^2} \log c_D \, ,
\label{S_eff_finiteN_gen}
\eeq
and $\tilde{\alpha}_k \equiv \alpha_k+\frac{1}{2}\delta_{k D}$.  Let
us first determine the region of the parameter space $ \{
\tilde{\alpha}_k \}$ where the partition function
eq. (\ref{finiteZ_gen}) exists. To that aim it is worthwhile to pass
to the eigenvalues $t_1,t_2,\ldots,t_D$ of $T$ in the integral
(\ref{finiteZ_gen}), as we did in eq. (\ref{finiteZ}), thus obtaining
a $D$-dimensional integral. The condition which prevents there being a
singularity at the point where all the $t_\mu$'s are zero is
\beq
\label{cond1}
(D-1)+\frac{D(D-1)}{2}+N^2 \left( \sum_{k=1}^{D} k \tilde{\alpha}_k - \frac{D(D+1)}{2N^2} \right) >-1
\eeq 
as one can see by passing to high-dimensional polar
coordinates\footnote{The first term of eq. (\ref{cond1}) is the
contribution from the radial part of the polar measure, the second is
from the Vandermonde, and the remaining terms are from the action. The
integral over the orthogonal group does not generate any
singularity.}. More generally, the integrand function does not have
singularities on the $p$-dimensional hyperplanes where $D-p$ of the
variables $t_\mu$'s are zero if and only if
\beq
\label{condD}
(D-p-1)+\frac{(D-p)(D-p-1)}{2}+N^2 \left( \sum_{k=p+1}^{D} (k-p)
\tilde{\alpha}_k - \frac{(D-p) (D+1)}{2 N^2} \right)>-1 \, , 
\eeq
for $0 \leq p \leq D-1$. In the large $N$ limit, the conditions in
eq. (\ref{condD}) relax to
\beq
\label{condD_largeN}
\sum_{k=p+1}^{D} (k-p) \tilde{\alpha}_k \geq 0\, ,  \quad p=0,\ldots,D-1 \, . 
\eeq 
In particular note that $\tilde{\alpha}_D \geq 0$. We call ${\cal D}$
the region in the parameter space $\{ \tilde{\alpha}_k \}$ which is
determined by the conditions in eq. (\ref{condD_largeN}), and from  
now on we shall consider only values of the parameters $\{
\tilde{\alpha}_k \}$ which belong to ${\cal D}$. Obviously, this is a
natural generalization of the analogous region obtained in
eq. (\ref{domaintrue}).

The generalized action eq. (\ref{S_eff_finiteN_gen}) at
large $N$ reads
\beq
S[T,\alpha,\Lambda] \equiv \tr T \Lambda - \sum_{k=1}^{D}
\tilde{\alpha}_k \log c_k \, ,
\label{S_effgen}
\eeq
and in the same limit the main contribution to the partition function
(\ref{finiteZ_gen}) comes from the global minima of $S$. Such
minima can be in the interior of the integration region or on
the boundaries of it.  In the former case, the saddle-point
equations are
\beq
\label{saddle_gen}
\frac{\partial}{\partial_{T_{\mu \geq \nu}}} S[T,\alpha,\Lambda]= \lambda_{\mu} \delta_{\mu\nu}-
\sum_{k=1}^{D} \tilde{\alpha}_k \frac{1}{c_k}\frac{\partial c_k}{\partial T_{\mu \geq \nu}}=0 \, .
\eeq
Any matrix $T$ which is a solution of eq.~(\ref{saddle_gen}) must be
diagonal. In fact, taking the commutator of eq.~(\ref{saddle_gen})
with $T$ yields $[\Lambda,T]=0$ because $T$ commutes with any other
function of $T$. Writing the commutator in components reads
$(\lambda_\mu-\lambda_\nu) T_{\mu\nu}=0$, i.e. $T$ is diagonal. Thus,
letting $T= \delta_{\mu \nu} t_\mu$, the saddle-point equations are
equivalent to the following system of non-linear algebraic equations
\beq
\label{system1}
\lambda_\mu=\sum_{k=1}^{D} \tilde{\alpha}_k \frac{1}{c_k}\frac{\partial c_k}{\partial t_{\mu}} \, , \quad \mu=1,\ldots,D \, .
\eeq
The case where the absolute minima of the action $S$ are instead on
the boundary of the integration region can occur only if some
parameters $\tilde{\alpha}_k$ are identically zero. In fact, if all the
parameters $\tilde{\alpha}_k$ are different from zero, then the action is
positively divergent when at least one $t_\mu$ is zero, and thus there
cannot be any minima on the boundary.  

For the moment, let us restrict the discussion to the case where all
the parameters $\tilde{\alpha}_k$ are strictly positive. We call
${\cal D}^{+} \subset {\cal D}$ such a region of the parameter space.
It is straightforward then to show that in ${\cal D}^{+}$ the system
in eq. (\ref{system1}) has only one real positive solution (that is a
set of $\{t_1>0, \ldots t_D>0 \}$ which fulfills eq. (\ref{system1})),
and it is actually the single global minimum of
eq. (\ref{S_effgen}). In fact, the linear combination $\tr T
\Lambda=\sum_\mu \lambda_\mu t_\mu$ and all the elementary symmetric
functions $c_k$ are multilinear ($k$-affine) functions in the
variables $t_1,\ldots,t_D$, as one can directly see from the
definition (\ref{c_definition}). As such they are convex functions. Also the
function $-\log(x)$ is convex for $x>0$, and therefore the action $S$
in eq. (\ref{S_effgen}) is a convex function, being a finite linear
combination with positive coefficients of convex functions.  Moreover,
we show that $S$ is also bounded from below.  In fact, we can prove it
by using the following inequality
\beq
\label{inequ}
\sum_{k=1}^D \tilde{\alpha}_k \log c_k \leq \sum_{k=1}^D k \tilde{\alpha}_k 
 \log c_1 \ \, .
\eeq
The proof of the inequality (\ref{inequ}) is by induction. For $D=1$
it is an identity. Let us suppose that eq. (\ref{inequ}) is valid for
$D-1$. Therefore we have
\beqa
\sum_{k=1}^D \tilde{\alpha}_k \log c_k  &\leq& 
 \sum_{k=1}^{D-1} k \tilde{\alpha}_k \log c_1+ \tilde{\alpha}_D \log
 \left( \frac{c_D}{c_{D-1}} \frac{c_{D-1}}{c_{D-2}} \cdots
 \frac{c_1}{c_0} \right) \nonumber \\ &\leq& \sum_{k=1}^{D-1} k
 \tilde{\alpha}_k \log c_1+ \tilde{\alpha}_D \log \left(
 \frac{c_1}{c_0} \right)^D = \sum_{k=1}^{D} k
 \tilde{\alpha}_k \log c_1 \, ,
\eeqa
where we used repeatedly Newton's inequalities $c_k^2 \geq c_{k-1}
c_{k+1}$ for $1 \leq k \leq D-1$, in the form $c_{k+1}/c_k \leq
c_k/c_{k-1}$, and the fact that $\tilde{\alpha}_D$ is positive in
${\cal D}$.  By applying the inequality (\ref{inequ}) to the effective
action eq. (\ref{S_effgen}) we get
\beq
S[T,\alpha,\Lambda] \geq \lambda_1 c_1- \sum_{k=1}^{D} k \tilde{\alpha}_k \log c_1 \, ,
\eeq
because $\tr T \Lambda \geq \lambda_1 \sum_{\mu=1}^D t_{\mu}$. Since the function 
 $a x-b \log(x) \geq b(1- \log \frac{b}{a})$ for any $a,b,x$ real and positive, we
finally obtain a lower bound for the  action
\beq
\label{bound_Seff}
S[T,\alpha,\Lambda] \geq A \left ( 1-  \log \frac{A}{\lambda_1 } 
\right) \, ,
\eeq
where $ A \equiv \sum_{k=1}^{D} k \tilde{\alpha}_k $ is positive in
${\cal D}$, as follows from eq. (\ref{condD_largeN}) with
$p=0$.\footnote{Note that the lower bound in eq. (\ref{bound_Seff}) is
actually valid everywhere in ${\cal D}$, and not only for $\{
\tilde{\alpha}_k \} \in {\cal D}^{+}$ as our proof does not rely on
such a restrictive hypothesis.}

All the above shows that when $\{ \tilde{\alpha}_k \} \in {\cal
D}^{+}$ the action $S$ is continuous, lower bounded and convex in the
integration region. From the additional observation that the action is
linearly divergent when any $t_\mu$ is large and logarithmically
divergent when any $t_\mu$ is close to zero we conclude that
necessarily the action has one and only one global minimum, and it
must be in the region $t_\mu>0$, $\forall \mu$.  We call such a
minimum $\bar{t} \equiv \{ \bar{t}_1,\ldots, \bar{t}_D\}$, 
$\bar{t}_\mu>0$. 

The large $N$ limit of the model is controlled by the behaviour of
$\bar{t}$ as a function of $\tilde{\alpha}_k$. In the following we
enumerate a series of properties of $\bar{t}$. To that aim is
worthwhile to recall two useful properties of the elementary symmetric
functions \cite{MacDonald,Niculescu,Chatur}. First, the $k$-th order
symmetric function $c_k$ can always be decomposed as the sum of a
$t_\mu$-dependent part and a $t_\mu$-independent part:
\beq
\label{recurr}
c_k=t_\mu c_{k-1}^{(\mu)}+ c_k^{(\mu)} \, ,
\eeq
where we defined $c_{k}^{(\mu)} \equiv \left. c_{k}
\right|_{t_\mu=0}$, i.e. the $k$-th elementary symmetric function of
$\{ t_1,t_2,\ldots,t_D \}$ omitting $t_\mu$. Note that $\partial_\mu
c_k=c_{k-1}^{(\mu)}$. Second, the following equality holds
\beq
\label{property2}
\sum_{\mu=1}^{D} t_\mu c_{k}^{(\mu)}=c_{k+1} \, , \quad k=0,\ldots,D-1 \, .
\eeq
Let us see now what consequences these properties have on $\bar{t}$. 
\begin{enumerate}
\item The solution $\bar{t}$ of eq. (\ref{system1}) is upper bounded by 
\[
 \bar{t}_\mu \lambda_\mu   = \sum_{k=1}^{D}  \tilde{\alpha}_k  
\frac{\bar{t}_\mu c_{k-1}^{(\mu)}}{c_k} \leq \sum_{k=1}^{D}  \tilde{\alpha}_k  
\, , \quad \forall \mu=1,\ldots,D \, ,
\]
because from eq. (\ref{recurr}) $c_k \geq t_\mu c_{k-1}^{(\mu)}$.
\item The solution $\bar{t}$ of eq. (\ref{system1}) is lower bounded by
\beq
\bar{t}_\mu = \frac{1}{\lambda_\mu} \sum_{k=1}^{D}  
\tilde{\alpha}_k \frac{\bar{t}_\mu c_{k-1}^{(\mu)}}{c_k}
\geq \frac{\tilde{\alpha}_D}{\lambda_\mu}
\label{lowerbound}
\eeq
because all the terms in the sum are non negative and $\bar{t}_\mu
c_{D-1}^{\mu}/c_D=1$. Therefore $\tilde{\alpha}_D$ {\em has} to go to
zero for $t_\mu \to 0$. Note that this condition means that when
$\tilde{\alpha}_D > 0$ there cannot be any spontaneous symmetry
breaking at all, since none of the eigenvalues is vanishing. In other
words, if there is a phase transition, it must be on the plane
$\tilde{\alpha}_D=0$. 
\item The minima $\bar{t}_\mu$ are in general monotonic with respect to
 $\mu$. Subtracting two equations of the system (\ref{system1}) gives
\beq
\lambda_\mu-\lambda_\nu= (\bar{t}_\nu-\bar{t}_\mu) \sum_{k=2}^{D} \tilde{\alpha}_k 
\frac{c_{k-2}^{(\mu,\nu)}}{c_k}  \, ,
\label{diff_of_eq}
\eeq
and then the ordering $\lambda_1<\lambda_2<\ldots<\lambda_D$ implies
$\bar{t}_1>\bar{t}_2>\ldots>\bar{t}_D$.  On the other hand , from
eq. (\ref{diff_of_eq}) follows also $\bar{t}_\mu =
\bar{t}_\nu$ if and only if $\lambda_\mu = \lambda_\nu$  i.e. when
 the symmetry breaking terms are removed. We deduce that at any point
 of the region ${\cal D}^{+}$, the dimensionality of the system is
 $d=D$ and the original $O(D)$ symmetry is fully preserved. In this
 case we obtain (with all $\lambda_\mu=1$)
\beq
\langle t_\mu \rangle= \frac{{\cal A}}
{D} \, , \quad {\cal F}={\cal A}  \left(
 1- \log {\cal A} \right)-  \sum_{k=1}^{D} \tilde{\alpha}_k \log \left[
\frac{1}{D^k} 
\left(
\begin{array}{c}
D \\
k 
\end{array}
\right)
\right] \, ,
\label{Free_gen}
\eeq
with ${\cal A} \equiv \sum_{k=1}^{D} k \tilde{\alpha}_k$. 

\item Let us consider now the limit $\tilde{\alpha}_D \to 0$, while 
 keeping all the other $\tilde{\alpha}_{k< D}$ fixed. In this limit
 the free energy has to be continuous either there is a symmetry
 breaking or is not. Its limiting value is given by eq. (\ref{Free_gen})
 just with $\tilde{\alpha}_D$ set to zero everywhere, i.e.
\beq
{\cal F}_{D}={\cal A}'  \left(
 1- \log {\cal A}' \right)-  \sum_{k=1}^{D-1} \tilde{\alpha}_k \log \left[
\frac{1}{D^k} 
\left(
\begin{array}{c}
D \\
k
\end{array}
\right)
\right] \, ,
\eeq
with ${\cal A}' \equiv \sum_{k=1}^{D-1} k \tilde{\alpha}_k$.
If there is symmetry breaking  then $\bar{t}_D \to 0$ (it is the smallest
 eigenvalue) but no other $\bar{t}_\mu$ can go to zero. This is because
 if there are at least two $\bar{t}_{D-1}, \bar{t}_{D} \to 0$ then
 $c_{D-1} \to 0$ and eq. (\ref{system1}) would be 
inconsistent in the limit (l.h.s. is finite whereas r.h.s. is infinite). 
The free energy for a $(D-1)$-dimensional broken phase (with $\alpha_D=0$)
would be:
\beq
{\cal F}_{D-1}={\cal A}'  \left(
 1- \log {\cal A}' \right)-  \sum_{k=1}^{D-1} \tilde{\alpha}_k \log \left[
\frac{1}{(D-1)^k} 
\left(
\begin{array}{c}
D-1 \\
k
\end{array}
\right)
\right] \, .
\eeq
In general ${\cal F}_{D} \leq {\cal F}_{D-1}$ with the equality only for
$D=2$, or $D>2$ and
$\tilde{\alpha}_2=\ldots=\tilde{\alpha}_{D-1}=0$. We conclude that
there is not spontaneous symmetry breaking when $\tilde{\alpha}_D \to
0$, unless for $D=2$, or $D>2$ and
$\tilde{\alpha}_2=\ldots=\tilde{\alpha}_{D-1}=0$ (which is actually
the case we considered in Section \ref{modelSec}). 
\end{enumerate}

It remains to consider the ``wedge'' region ${\cal D}/{\cal D}^{+}$ of
the phase space, where some of the $\tilde{\alpha}_k$ are negative.
In this case the action $S$ is no longer a convex function, but it
still possible to prove that it has only one global minimum.  The
proof goes as follows. First of all, $S$ is still lower bounded by
the same bound as in eq. (\ref{bound_Seff}), and it is divergent
towards $+ \infty$ at the boundaries of the integration region, hence
it must have at least one local minimum. Secondly, if $S$ has more
than one local minimum then the system of equations (\ref{system1})
would have multiple solutions $\bar{t}_\mu$ for a set of values of the
parameters $\{
\tilde{\alpha}_k
\}$. We know already that when $\{ \tilde{\alpha}_k \}$ is in ${\cal D}^{+}$
the solution is unique, therefore there must exists a value $\{
\tilde{\alpha}_k' \}$ of the parameters where multiple solutions merge
together into the unique one. This implies that the Jacobian $\det
\partial
\bar{t}(\tilde{\alpha}) / \partial \tilde{\alpha}$ has to be singular or zero
 for that particular value of $\tilde{\alpha}'$. However we show now that this is not
possible. In fact, let us write the system of equations
(\ref{system1}) in the more compact form:
\beq
\label{compact}
\lambda=G[t(\tilde{\alpha})] \cdot \tilde{\alpha}
\eeq
where $\lambda=(\lambda_1,\ldots,\lambda_D)$, $ G_{\mu k}[t] \equiv
c_{k-1}^{(\mu)}/c_k$ and
$\tilde{\alpha}=(\tilde{\alpha}_1,\ldots,\tilde{\alpha}_D)$.  Equation
(\ref{compact}) implicitly defines the vector function
$t(\tilde{\alpha})$ as a function of $\tilde{\alpha}$.  We take the
total derivative of the component $\mu$ of eq. (\ref{compact})
w.r.t. $\tilde{\alpha}_i$ and compute the determinant w.r.t the
indexes $\mu,i$ of the obtained expression. One has:
\beq
\det_{\mu \sigma} \frac{ (G[t(\tilde{\alpha})]  \cdot 
\tilde{\alpha} )_\mu}{\partial t_{\sigma}} \, \, 
\det_{\sigma i} \frac{\partial t_{\sigma}(\tilde{\alpha})}{\partial \tilde{\alpha}_i} =
(-1)^{D} \det G[t(\tilde{\alpha})] \, .
\label{detdet}
\eeq
The first determinant on the l.h.s. of eq. (\ref{detdet}) is regular
and not zero at $\{ \tilde{\alpha}' \}$, otherwise the
eq. (\ref{system1}) would not admit any implicit solution but we know
it must exist (because the existence of a global minimum).  The
determinant on the r.h.s. is $\det G= \Delta(t)/ \prod_{j=1}^{D} c_j$,
where $\Delta(t)$ is the Vandermonde determinant. This expression is
finite and it zero only if at least two eigenvalues $t_\mu, t_\nu$ are
equal each other and then, by means of eq. (\ref{system1}), it must be
$\lambda_\mu = \lambda_\nu$ which is not possible by hypothesis.  This
ends the proof that for any given $\{
\tilde{\alpha}_k \}$ in the ``wedge'' regions the action has
only one local minimum in the interior of ${\cal D}$, which is then
also a global one. The qualitative behaviour of this critical point as 
a function of the parameters $\tilde{\alpha}$ goes
as for the case in ${\cal D}^{+}$. After removing the symmetry breaking
terms $\lambda \to 1$, the critical point becomes completely symmetric
in its variables and it corresponds to an un-broken phase with  $O(D)$
symmetry.

\sect{Discussion and conclusions} \label{concSec}

In this paper we have introduced a multi-matrix model where the
Hermitian matrices $X_\mu$ are interacting through all the elementary
symmetric functions of the correlation matrix $T_{\mu \nu} = \Tr(X_\mu
X_\nu)/N $. The main reason for the choice of such a model relies in
its interesting features: first, it is manifestly $O(D) \times SU(N)$
invariant, and it allows the study of the issue of the spontaneous
symmetry breaking of $O(D)$ symmetry in the large-$N$ limit.
Secondly, the action of the model is real, positive definite and it
does not contain any Grassmann variables. This is most useful for
understanding what we can actually expect from a model without a
complex action or rapidly fluctuating potentials.  Understanding the
effect of a complex action, which is a notorious difficult problem,
requires also realizing first what could happen when it is not there.
Third, it allows a number of possible ``degenerate configurations'' in
the matrix integration measure and our aim is to understand their
r\^{o}le in a scenario of spontaneous symmetry breaking. Finally, the
model is considerably simple and can be solved analytically, being the
interaction among the matrices only through the $O(D)$ ``spatial''
symmetry and not through the SU($N$) ``internal'' symmetry (for which
there is just a Gaussian weight). We introduced a number of parameters
which allows to tune the relative weight of the elementary symmetric
functions of the model, and then we focused our attention on the
phases of the model in the space of the parameters when $N$ is large.
This has been done in two steps: first in Section
\ref{modelSec} by studying in full detail a simple case where only two
symmetric functions are ``switched on'' (the trace and the
determinant), and afterwards in Section \ref{genSec} by considering
the more general case where all the symmetric functions are present at
the same time. In both cases we found that the $O(D)$ symmetry is
broken only in the limit $\alpha_D \to -1/2$ for $D=2$ or for $D>2$
and $\alpha_2=\ldots=\alpha_{D-1}=0$. In these cases the
dimensionality of the model collapses down to one dimension.

The qualitative explanation of such a behaviour is simple.  Degenerate
configurations of the matrices such that the correlation function
$T_{\mu\nu}$ has zero eigenvalues, dominate the matrix integration in
the large $N$ limit, when the parameters of the model are tuned to a
critical value. In particular the parameter $\alpha_D$ (which is
coupled to the determinant, i.e. the elementary symmetric function
most sensitive to ``degenerate'' configurations) is to be tuned to the
critical value $\alpha_D=-1/2$ for compensating an analogous
``centrifugal'' term coming from the Jacobian (see Appendix). At that
precise value of $\alpha_D=-1/2$, the measure collapse down to one
dimensional configurations, quite independently from the presence of
other symmetric functions but the trace. This is most evident from the
explicit solutions in Sections
\ref{modelSec}.

The symmetry breaking mechanism of the model in this paper is
therefore due to the existence of directions in the matrix integral
along which the measure is identically zero. These directions are
where the matrices are linearly dependent, with different degree of
degeneracy. We learned also that the reality of the action does not
seem to stop a generic Hermitian multi-matrix models with $O(D) \times
SU(N)$ symmetry from having a spontaneous symmetry breaking of $O(D)$
symmetry when $N$ is large.  Of course this does not prevent other
real-action multi-matrix models having different patterns of
spontaneous symmetry breaking, nor does not say anything about the
r\^ole played by a possible complex term in the action. For all these
reasons our findings do not contradict the analysis of
\cite{largeN11,largeN12,largeN13,largeN14,largeN21,largeN22,NV,Hotta:1998en,review1}. 
It would be interesting to carry out the analysis contained in this
paper to an extended version of the model where the coupling constants
$\tilde{\alpha}$ are allowed to be complex numbers. The action would
be complex then, and a different pattern of symmetry breaking seems to
be possible.\\ There are extensions of the model where the matrices
are not Hermitian but real symmetric or symplectic. The only changes
are in slightly different factors in the Jacobian (see Appendix) and
they do not affect the large $N$ results of this paper which still
would hold in those generalized cases. We conclude by observing that
the reason why we can solve this multi-matrix model is that the
interaction among the matrices is only through the correlation matrix
$T_{\mu \nu}$. For the rest the matrices are actually not interacting
with the full internal $SU(N)$ symmetry group, the interaction being
just a Gaussian factor. In fact adding a quartic or higher order term
to the action (i.e. terms like $\Tr X_\mu X_\nu X_\mu X_\nu$ and $\Tr
X_\mu^2 X_\nu^2$) would probably change drastically this scenario, but
it would also be more difficult to solve, as happens for multi-matrix
models like the Yang-Mills integrals.

\vspace{.5cm}

\noindent
\underline{Acknowledgments}: 

The authors would like to thank P.Austing, G. Cicuta, T.~Jonsson,
C.F.Kristjansen, for useful discussions. We also acknowledge
J. Nishimura for collaborating at an early stage of this work.  The
work of G.V. is supported by the EU network on ``Discrete Random
Geometry'', grant HPRN-CT-1999-00161.

\begin{appendix}
\sect{Appendix} \label{Appendix}

For the sake of readability, in this Appendix we compute the Jacobian
of the transformation in eq.~(\ref{ZT}).  It is a well-known result
which has appeared several times in the literature, e.g.
\cite{David:1992vv,YF,Ing,Sie,CMV}. The technique we use here is similar 
to the one in \cite{David:1992vv}. The integral in eq.~(\ref{Z}) is of the form
\beq
\label{I}
I \equiv \int \prod_{\mu=1}^D \dd X_\mu \, f(T[X])  \, 
\eeq
where $f$ is a real function and the $U(N)$-invariant integration
measure for each Hermitian matrix is $dX=\prod_{i=1}^{N} dX_{ii}
\prod_{i>j} d \mbox{Re} X_{ij} \, d \mbox{Im} X_{ij}$ as usual. First,
by inserting the definition eq.~(\ref{matrixT}) of the matrix $T$ in the
formula (\ref{I}) by means of Dirac $\delta$-functions, we can
equivalently write
\[
I=\int_{T \geq 0} dT \, f(T) J(T) \, , \qquad J(T) \equiv \int
\prod_{\mu=1}^D \dd X_{\mu} \prod_{\alpha \geq \beta}^D \delta(
T_{\alpha \beta}- \frac{1}{N}\Tr{X_{\alpha} X_{\beta}} ) \, .
\]
The Jacobian $J(T)$ can be evaluated by using the integral
representation of the $\delta$-function
\beqa
J(T)&=& \int \prod_{\mu=1}^D \dd X_\mu \prod_{\alpha \geq \beta}^D \int
\frac{d \Omega_{\alpha \beta}}{2 \pi} e^{ i \Omega_{\alpha \beta}
(T_{\alpha\beta}-\frac{1}{N}\Tr{X_\alpha X_\beta} )}
\nonumber \\
 &=& \widetilde{{\cal C}}_{N,D} \int d \Omega \, \frac{e^{i \, \tr
 \Omega T}}{\det (i\Omega)^{N^2/2}} \, \qquad
\widetilde{{\cal C}}_{N,D}= \frac{N^{\frac{D N^2}{2}} 
\pi^{\frac{D}{2}(N^2-D-1)}}{2^{D \left[ \frac{N(N-1)}{2}+1 \right] }} \, , 
\eeqa
where in the last equation we performed the Gaussian integral over the
matrices $X_\mu$, and we collected the elements $\Omega_{\mu \nu}$
into a real symmetric matrix $\Omega$ (giving an additional factor
from the measure). The real-symmetric matrix $T$ can be diagonalized
by an orthogonal matrix $O$, i.e. $T=O t O^{T}$ where $t$
is a diagonal matrix, with diagonal elements $t_\mu \geq 0$. 
We change the matrix variables $\Omega \to W$ where $W=O^{T}
\Omega O$ and we have $dW=d\Omega$, $\det(i W)=\det(i \Omega)$ and
\beqa
J(T)&= &\widetilde{{\cal C}}_{N,D} \int dW \frac{e^{ i \tr W t }}{\det
(i W)^{N^2/2}} \nonumber \\ &=& \widetilde{{\cal C}}_{N,D} (\det
T)^{\frac{N^2-D-1}{2}} \int dW \frac{e^{ i \tr W}}{[\det i W]^{N^2/2}}
\eeqa
where in the last equation we apply the transformation $W_{\mu\nu}
\rightarrow W_{\mu\nu}/ \sqrt{t_\mu t_\nu}$. The remaining
$T$-independent integral is completely factorized and it is equal to
$2^D \pi^{\frac{D(D+3)}{4}}/\prod_{k=1}^D \Gamma(\frac{N^2-k+1}{2})$.
Finally we obtain
\beq
J(T)= \frac{N^{\frac{DN^2}{2}}
\pi^{\frac{D}{4}(2N^2-D+1)}}{2^{D\frac{N(N-1)}{2}}\prod_{k=1}^D \Gamma
\left( \frac{N^2-k+1}{2} \right) } (\det T)^{\frac{N^2-D-1}{2}} \, .
\eeq
The results in this appendix are valid for $N^2 \geq D$, which is fine
for the large $N$ analysis of this paper.
\end{appendix}

\indent

\end{document}